\documentclass[12pt,onecolumn,showpacs,amsmath,amssymb]{revtex4}
\begin{document}

\vspace*{4cm}

\title{Nonlinear waves on the surface of a dielectric liquid
in a strong tangential electric field}

\author{N. M. Zubarev}
\email{nick@ami.uran.ru}

\affiliation{Institute of Electrophysics, Ural Branch, Russian
Academy of Sciences,\\ 106 Amundsen Street, 620016 Ekaterinburg, Russia\\
tel.: +7 343 2678776; fax: +7 343 2678794.}

\begin{abstract}

The nonlinear dynamics of the free surface of an ideal dielectric liquid
in a strong electric field is studied. The equation for the evolution of
surface electrohydrodynamic waves is derived in the approximation of
small surface-slope angles. It is established that the equation can be
solved for liquids with sufficiently high values of the permittivity. This
makes it possible to describe the interaction of the counter-propagating
waves.

\end{abstract}


\maketitle


\newpage

It is well known \cite{mel} that the tangential electric field, as
opposed to the normal electric field, has a stabilizing effect on
the free surface of a dielectric liquid. The dispersion relation for the
surface waves propagating along the electric field direction 
has the following form \cite{mel1,sh}:
$$
\omega^2=gk+\frac{E^2\beta(\varepsilon)}{4\pi\rho}\,k^2
+\frac{s}{\rho}\,k^3,
$$
where $k$ is the wave number, $\omega$ is the frequency, $g$ is the
acceleration of gravity, $s$ is the surface tension coefficient,
$\rho$ is the mass density, $E$ is the external electric field strength, 
$\beta=(\varepsilon-1)^2/(\varepsilon+1)$, and $\varepsilon$ is the
dielectric constant of a fluid (we use the cgs
electrostatic system of units). It is seen from this relation that, if the
electric field is sufficiently strong 
($E^2\gg\beta^{-1}\!\sqrt{gs\rho}$), 
the waves with wavelengths $\lambda=2\pi/k$ in the range
\begin{equation}
s E^{-2}\beta^{-1}\ll \lambda\ll E^2\beta(\rho g)^{-1}
\label{usl}
\end{equation}
will travel without dispersion, i.e., $\omega\propto k$. Their propagation
is described by the linear wave equation \cite{mel} 
\begin{equation}
\eta_{tt}=c^2\eta_{xx},
\label{lin}
\end{equation}
where $\eta(x,t)$ is the elevation of the free surface, and $c$ is the 
(constant) wave velosity. Its general solution is 
$$
\eta(x,t)=f(x+ct)+g(x-ct),
$$
where $f$ and $g$ are arbitrary functions that play the role of the
Riemann invariants (they are constant along the corresponding
characteristics). The problem of taking into account the system
nonlinearities amounts to studying the manner in which the Riemann
invariants of the linear wave equation will evolve under their influence. 
 
For the classical problem of long waves in shallow water, the
evolution of the surface is also described (in the linear approximation)
by the equation (\ref{lin}). As is known (see, for example,
Ref.~\cite{new}), the nonlinearity determines the tendency for the
formation of singularities. Such tendency can be suppressed by the
linear dispersion influence. The asymptotic behaviour of the (adiabatic)
Riemann invariants is then described by the Korteweg--de Vries equation.
Below we will demonstrate that, when investigating the nonlinear waves on
the surface of a deep fluid layer subjected to a strong tangential
electric field, we encounter the situation of a fundamentally different
kind. The point is that, for liquids with high $\varepsilon$ values,
the Riemann invariants of the linear wave equation, taken separately, are
not distorted by nonlinearities. Of course, this is not to say that 
the nonlinearity influence can be neglected. The nonlinearity is
responsible for the interaction of the oppositely directed waves.

Note that the influence of the nonlinearity on the waves propagating on
the surface of a fluid in a tangential electric or magnetic field
(these problems are similar from the mathematical point of view) is
usually investigated in the quasi-monochromatic approximation (see
\cite{el,rol} and the references therein). This approach allows one to
obtain immediately a nonlinear Schr\"odinger equation for the complex
amplitude of travelling waves and, in so doing, to circumvent the
difficulties associated with the necessity of analysing the nonlinear
integro-differential equations that result from reducing the original
equations of motion to lower dimensional equations for the surface
evolution. In the present work it will be shown that, in the strong-field
limit, such integro-differential equations can be solved and,
consequently, the condition of quasi-monochromaticity can be replaced by
the weaker condition (\ref{usl}).

Consider the potential motion of an ideal liquid of infinite depth in an
external tangential electric field (the potential flow approximation holds
if the characteristic wavelength $\lambda$ far exceeds the penetration
depth of the vortex motion $\sqrt{\nu/\omega}$, where $\nu$ is the kinematic
viscosity of the fluid \cite{lanli}). We assume that, in the unperturbed
state, the boundary of the liquid is a flat horizontal surface 
$z=0$ (the $z$ axis is normal to the surface of the liquid), and the
electric field vector is directed along the $x$ axis of the 
Cartesian coordinate system. The function $\eta(x,t)$ specifies the
deviation of the boundary from the plane, i.e., the liquid occupies
the region $z<\eta(x,t)$.

The velosity potential for an incompressible liquid $\Phi$
satisfies the Laplace equation
$$ \nabla^2\Phi=0 $$
with the following conditions at the boundary and at infinity:
$$
\Phi_t+\frac{(\nabla\Phi)^2}{2}= \frac{(\varepsilon-
1)(\nabla\varphi\cdot\nabla\varphi'-E^2)}{8\pi\rho}, \qquad
z=\eta,
$$
$$
\Phi\to 0, \qquad z\to-\infty,
$$
where $\varphi$ and $\varphi'$ are the electric-field potentials in and
above the liquid. The expression on the right-hand side of the dynamic
boundary condition (nonstationary Bernoulli equation) is responsible for
the electrostatic pressure at the interface between an ideal dielectric
liquid and air (or vacuum) in the absence of free electric charges
\cite{lanlif}. The evolution of the free surface is determined by the
kinematic relation: 
$$
\eta_t=\Phi_z-\eta_x\,\Phi_x, \qquad z=\eta.
$$

The electric potentials $\varphi$ and $\varphi'$ satisfy the Laplace
equations 
$$
\nabla^2\varphi=0, \qquad \nabla^2\varphi'=0.
$$
Since the electric field potential and normal component of the
displacement vector have to be continuous at the interface, we should add
the following conditions at the boundary:
$$ 
\varphi=\varphi', \qquad z=\eta,
$$
$$
\varepsilon\partial_n \varphi=\partial_n\varphi', \qquad
z=\eta,
$$
where $\partial_n$ denotes the derivative along the normal to the surface
$z=\eta$. The system of equations is closed by the condition of the
electric field uniformity at an infinite distance from the surface:
$$
\varphi'\to -Ex, \qquad z\to+\infty,
$$
$$
\varphi\to -Ex,
\qquad z\to-\infty.
$$

The above-written equations of motion have a Hamiltonian structure, and
the functions $\eta(x,t)$ and $\psi(x,t)=\Phi|_{z=\eta}$ are canonically
conjugated quantities \cite{zah,kuz}:
$$
\psi_t=-\frac{\delta H}{\delta\eta},
\qquad
\eta_t=\frac{\delta H}{\delta\psi},
$$
where the Hamiltonian $H$ coincides with the total energy of the system:
$$
H=\int\limits_{z\leq\eta}\frac{(\nabla\Phi)^2}{2} dx\,dz-
\int\limits_{z\geq\eta}\frac{(\nabla\varphi')^2-E^2}{8\pi\rho}dx\,dz-
\varepsilon\int\limits_{z\leq\eta}\frac{(\nabla\varphi)^2-E^2}
{8\pi\rho}dx\,dz.
$$

It is convenient to introduce perturbations of the electric-field
potentials: 
$$
\tilde\varphi'=\varphi'+Ex, \qquad \tilde\varphi=\varphi+Ex.
$$
It can be readily shown that the perturbed potentials satisfy
the Laplace equations with trivial conditions at infinity,
$$
\tilde\varphi'\to 0, \qquad z\to+\infty, 
$$
$$
\tilde\varphi\to 0, \qquad z\to-\infty, 
$$
and the following conditions at the boundary:
\begin{equation}
\tilde\varphi'=\tilde\varphi, \qquad z=\eta,
\label{gra}
\end{equation}
\begin{equation} \varepsilon\left(\tilde\varphi_z-
\eta_x\,\tilde\varphi_x\right) =\tilde\varphi'_z-
\eta_x\,\tilde\varphi'_x
-E(\varepsilon-1)\eta_x, \qquad z=\eta.
\label{gran}
\end{equation}
Substituting the perturbed potentials into the Hamiltonian and
then integrating it by parts over $z$, we obtain:
\begin{equation}
H=\int\left[
\frac{\psi\left.(\Phi_z-\eta_x\Phi_x)\right|_{z=\eta}}{2}-
\frac{E(\varepsilon-
1)\phi\eta_x}{8\pi\rho}\right]dx,
\label{gam}
\end{equation}
where we put
$\phi(x,t)=\tilde\varphi|_{z=\eta}=\tilde\varphi'|_{z=\eta}$.

Our next problem is to express the integrand in (\ref{gam}) in terms
of the functions $\eta$ and $\psi$. Let the characteristic angles of the
surface slope be small, $\alpha\sim|\eta_x|\ll 1$ (the boundary
perturbation amplitude is much less than the characteristic wavelength). 
Then the derivatives of the potentials involved in (\ref{gran}) and
(\ref{gam}) can be expanded into a power series:
\begin{equation}
\Phi_x|_{z=\eta}=\hat T_{+}\partial_x\hat T_{+}^{-1}\psi, \qquad
\Phi_z|_{z=\eta}=-\hat T_{+}\partial_x\hat H\hat T_{+}^{-1}\psi,
\label{r1}
\end{equation}
\begin{equation}
\tilde\varphi_x|_{z=\eta}=\hat T_{+}\partial_x\hat T_{+}^{-1}\phi,
\qquad \tilde\varphi_z|_{z=\eta}=-
\hat T_{+}\partial_x\hat H\hat T_{+}^{-1}\phi,
\label{r2}
\end{equation}
\begin{equation}
\tilde\varphi'_x|_{z=\eta}=\hat T_{-}\partial_x\hat T_{-}^{-1}\phi,
\qquad \tilde\varphi'_z|_{z=\eta}=
\hat T_{-}\partial_x\hat H\hat T_{-}^{-1}\phi,
\label{r3}
\end{equation}
where $\hat T_{\pm}=e^{\mp\eta\partial_x\hat H}$, and $\hat H$
is the Hilbert transform:
$$
\hat{H}f(x)=\pi^{-1}\!\!\int\!f(x')(x'-x)^{-1}dx'.
$$
In deriving these relationships we have taken into account that
$\left.f_z\right|_{z=0}=\pm\hat{H}\!\left.f_x\right|_{z=0}$ 
and $f|_{z=\eta}=\hat{T}_{\mp}(f|_{z=0})$ for harmonic functions
that decay as $z\to\pm\infty$.

Inserting (\ref{r2}) and (\ref{r3}) into the conditions (\ref{gra}) and
(\ref{gran}) and using the method of successive approximations, we obtain
the expression for $\phi$, correct to second order in small quantities:
$$
\phi\approx-\gamma E\hat H\eta-
\gamma^2E\eta\eta_x-\gamma^2E\hat H(\eta\hat H\eta_x),
$$
where we have introduced the notation
$\gamma=(\varepsilon-1)/(\varepsilon+1)$. If we substitute it together
with (\ref{r1}) into the Hamiltonian and restrict ourselves to
second- and third-order terms, we finally obtain:
$$
H=-\frac{1}{2}\int\left[\psi\hat H\psi_x+
\eta\left((\hat H\psi_x)^2-\psi_x^2\right)\right] dx-
\frac{E^2\beta}{8\pi\rho}\int\left[\eta\hat H\eta_x+
\gamma\eta\left((\hat H\eta_x)^2-\eta_x^2\right)\right] dx.
$$
The equations corresponding to this Hamiltonian are
\begin{equation}
\psi_t-\hat H\eta_x=\alpha\gamma\hat H(\eta\hat H\eta_x)_x+
\alpha\gamma(\eta\eta_x)_x+
\frac{\alpha}{2}\left[\gamma(\hat H\eta_x)^2-
\gamma\eta_x^2+(\hat H\psi_x)^2-
\psi_x^2\right]+O(\alpha^2),
\label{ur1}
\end{equation}
\begin{equation}
\eta_t+\hat H\psi_x=
-\alpha\hat H(\eta\hat H\psi_x)_x-
\alpha(\eta\psi_x)_x+O(\alpha^2),
\label{ur2}
\end{equation}
where we have passed to dimensionless variables using the
transformations 
$$
\psi\to\alpha\lambda E\beta^{1/2}(4\pi\rho)^{-1/2}\,\psi,
\qquad
\eta\to\alpha\lambda\,\eta,
$$
$$
t\to \lambda E^{-1}\beta^{-1/2}(4\pi\rho)^{1/2}\,t, \qquad
x\to \lambda\, x.
$$
These nonlinear nonlocal equations describe the evolution of
small-amplitude surface waves. 

Applying the operator $\partial_x\hat{H}$ to Eq.~(\ref{ur1}),
differentiating Eq.~(\ref{ur2}) with respect to $t$, and then
subtracting the resulting expressions, we arrive at
\begin{equation}
\eta_{tt}-\eta_{xx}=\alpha \partial_x\hat{H}\!
\left[(\hat{H}\eta_t)^2-\gamma(\hat{H}\eta_x)^2-
(\gamma-1)\left(\eta\eta_{xx}+\hat{H}(\eta\hat{H}\eta_{xx})\right)
\right]+O(\alpha^2),
\label{ur3}
\end{equation}
where we have used the well-known relations
$2\hat{H}(f\hat{H}f)=(\hat{H}f)^2-f^2$ and $\hat{H}^{2}f=-f$. 
The function $\psi$ was eliminated from the right-hand side of this
equation with the help of the expression $\psi_x=\hat{H}\eta_t+O(\alpha)$,
which follows from (\ref{ur2}). One can see that in the linear
approximation, i.e., in the limit $\alpha\to 0$, the equation (\ref{ur3})
transforms into the linear wave equation of the form (\ref{lin}) and,
consequently, it describes the dispersionless wave propagation. 

Consider the influence of nonlinearity on
the waves that travel along or against the direction of the $x$ axis. We
will seek a solution of (\ref{ur3}) in the form 
$$
\eta(x,t)=\tilde\eta(y,\tau),
$$
where $y=x\pm t$ and $\tau=\alpha t$. In the leading order of the expansion
in the characteristic surface-slope angle $\alpha$,  
we obtain the following equation for the slow evolution of the wave: 
\begin{equation}
2\tilde\eta_{\tau}=\mp(\gamma-1)\partial_y\hat{H}\!
\left[(\hat{H}\tilde\eta_y)^2+
(\tilde\eta\tilde\eta_{yy})+
\hat{H}(\tilde\eta\hat{H}\tilde\eta_{yy})
\right]+O(\alpha),
\label{ur4}
\end{equation}
where the Hilbert transform operator acts with respect to the variable
$y$. Unfortunately, this nonlinear integro-differential equation cannot
be solved in the general case. One can assume that it describes the
formation of singularities in a finite time. This process may be
suppressed by the influence of linear dispersion which should be taken
into account for a finite-depth fluid layer.

Let us restrict our consideration to the degenerate case
\begin{equation}
\gamma(\varepsilon)-1=O(\alpha), \label{us1}
\end{equation}
when the equation (\ref{ur4}) becomes trivial:
$$
\tilde\eta_{\tau}=O(\alpha).
$$
This means that the quadratic nonlinearities have no effect on
the wave propagation, i.e., the nonlinearity does not distort the
Riemann invariants of the linear wave equation. 

It can easily be seen that the condition (\ref{us1}) holds for dielectric
liquids with sufficiently high values of the permittivity $\varepsilon$. 
Indeed, since $\gamma-1=-2(\varepsilon-1)^{-1}$, it can be rewritten as
\begin{equation}
\varepsilon\geq O(\alpha^{-1}).
\label{us2}
\end{equation}
Let, for example, $\alpha=0.1$, i.e., the wave amplitude be an order of
magnitude less than the characteristic wavelength, and the considered
small-angle approximation is quite applicable. Then the condition
(\ref{us2}), i.e., the condition $\varepsilon\geq O(10)$, is certainly
valid for such liquids as ethanol ($\varepsilon\approx 26$), 
nitrobenzol ($\varepsilon\approx 36$), and deionized water
($\varepsilon\approx 81$), which are often used in experiments.

Now consider the evolution of the free surface of a liquid with high
permittivity within the framework of the equation (\ref{ur3}) describing
the waves travelling in both possible directions. Putting $\gamma=1$ we,
up to terms of order $O(\alpha^2)$, obtain:
$$
\eta_{tt}-\eta_{xx}=\alpha\partial_x\hat{H}\!\left[(\hat{H}\eta_t)^2-
(\hat{H}\eta_x)^2\right].
$$
Applying the Hilbert transform operator to both parts of this equation and
introducing the auxiliary function $v(x,t)=\hat{H}\eta$, we find:
\begin{equation}
v_{tt}-v_{xx}=\alpha\partial_x(v_x^2-v_t^2).
\label{key}
\end{equation}
Thus, if the condition (\ref{us2}) holds true, it is possible to pass
from the integro-differential equation (\ref{ur3}) to the partial
differential equation. 

It is apparent that the key equation of the paper, i.e., the equation
(\ref {key}), has the following particular solutions: 
\begin{equation}
v=F(\xi), \qquad \xi=x+t,
\label{resh1}
\end{equation}
\begin{equation}
v=G(\zeta), \qquad \zeta=x-t,
\label{resh2}
\end{equation}
where $F$ and $G$ are arbitrary functions. These solutions correspond to
the dispersionless propagation of nonlinear waves along (or against) the
direction of the $x$ axis. Let us study the interaction of these waves.

We will seek the general solution of the equation (\ref{key}) as a power
series in the small parameter $\alpha$: 
$$
v=A(\xi,\zeta)+\alpha B(\xi,\zeta)+O(\alpha^2).
$$
For leading order we obtain the linear wave equation
\begin{equation}
A_{\xi\zeta}=0,
\label{u1}
\end{equation}
and the equation 
\begin{equation}
B_{\xi\zeta}=-(A_{\xi}A_{\zeta})_{\xi}-(A_{\xi}A_{\zeta})_{\zeta}
\label{u2}
\end{equation}
for order $O(\alpha)$. It is meaningless to consider higher orders of the
expansion since it is beyond the accuracy of the model (\ref{key}). 
The equation (\ref {u1}) has the general (D'Alembert) solution:
$$
A=F(\xi)+G(\zeta).
$$
The solution of the inhomogeneous equation (\ref {u2}) is given by the
following expression:
$$
B=\tilde F(\xi)+\tilde G(\zeta)-F_{\xi}G-FG_{\zeta},
$$
where we can set $\tilde F=0$ and $\tilde G=0$ without loss of generality.
An important feature of the described scheme for solving the equation
(\ref{key}) is that the leading-order solution is the sum of 
the exact particular solutions (\ref{resh1}) and (\ref{resh2}). Otherwise,
such procedure would give rise to secular terms, i.e., the terms that grow
algebraically with $x$ or $t$. As a result, the conditions of
applicability of the perturbation theory would be violated for  
$t>\alpha^{-1}$ and $|x|>\alpha^{-1}$.

Thus, the solution of Eq.~(\ref{key}) has the following form:
$$
v(x,t)=F(x+t)+G(x-t)-\alpha(FG)_x+O(\alpha^2).
$$
Returning to the function $\eta$, we find that the surface evolution
is described by the expression:
\begin{equation}
\eta(x,t)=f(x+t)+g(x-t)+
\alpha\partial_x\hat{H}\!\left(\hat{H}f\cdot\hat{H}g\right)+O(\alpha^2),
\label{sol}
\end{equation}
where $f=-\hat{H}F$ and $g=-\hat{H}G$. It is obvious from this nonlinear 
superposition formula that the interacting counter-propagating solitary waves
preserve their initial shapes and phases.

Thus, we have described the interaction of the counter-propagating waves
on the surface of an ideal dielectric liquid in the presence of a strong
tangential electrical field. It turns out that, even in the dispersionless
case, the nonlinear evolution of the system does not result in the
formation of singularities (e.g. wave breaking) if the condition
(\ref{us2}) is satisfied. As a consequence, the interaction of nonlinear
waves does not lead to the energy transfer towards small spatial scales.

It should be noted that the results of the above investigation can
be extended to the case of a ferrofluid in a tangential magnetic
field. All one has to do is to replace the electric field $E$ by the
magnetic field $H$ and the permittivity $\varepsilon$ by the magnetic
permeability $\mu$. In so doing the condition (\ref{us2}) for the
applicability of the solution (\ref{sol}) turns into the inequality 
$\mu\geq O(\alpha^{-1})$, which can be valid for some ferrofluids.

An analogous approach can be applied to describe the wave propagation at
the interface of two liquids with different permittivities $\varepsilon_1$
and $\varepsilon_2$. In the limiting case when the Atwood number
$A=(\rho_1-\rho_2)/(\rho_1+\rho_2)$ tends to unity (then the
upper fluid has no influence on the boundary motion), all the above
results remain valid with the only correction that $\varepsilon$ is now
the permittivity ratio, $\varepsilon=\varepsilon_1/\varepsilon_2$.  

\medskip

This study was supported by the ``Dynasty'' Foundation and the
International Center for Fundamental Physics in Moscow.


\begin{thebibliography}{}

\bibitem{mel}
J.R. Melcher, Field-Coupled Surface Waves, MIT Press,
Cambridge, MA, 1963.

\bibitem{mel1}
J.R. Melcher, Phys. Fluids 4 (1961) 1348.

\bibitem{sh}
M.I. Shliomos, Usp. Fiz. Nauk 112 (1974) 427 [Sov. Phys. Usp. 17 (1974)
153]. 

\bibitem{new}
A.C. Newell, Solitons in Mathematics and Physics, CBMS-NSF Regional
Conference Series in Applied Mathematics, Vol. 48, SIAM, Philadelphia, PA,
1985. 

\bibitem{el}
M.F. El-Sayed, D.K. Callebaut, Physica A 269 (1999) 235.

\bibitem{rol}
D.K. Rollins, B.K. Shivamoggi, Phys. Plasmas  8 (2001) 2930.

\bibitem{lanli}
L.D. Landau and E.M. Lifshitz, Course of Theoretical Physics, Vol. 6:
Fluid Mechanics, Nauka, Moscow, 1988; Pergamon, New York, 1987.

\bibitem{lanlif}
L.D. Landau and E.M. Lifshitz, Course of Theoretical Physics, Vol. 8:
Electrodynamics of Continuous Media, Nauka, Moscow, 1982; Pergamon, New
York, 1984.

\bibitem{zah}
V.E. Zakharov, Prikl. Mekh. Tekh. Fiz., No. 2 (1968) 86.

\bibitem{kuz}
E.A. Kuznetsov and M.D. Spector, Zh. \'Eksp. Teor. Fiz. 71 (1976) 262 
[Sov. Phys. JETP 44 (1976) 136].



\end{thebibliography}
\end{document}